\documentclass[12pt, oneside]{article}
\usepackage{jheppub}
\usepackage{graphicx}
\usepackage{amsmath}								
\usepackage{color}
\usepackage{mathtools}
\usepackage{amssymb}

\usepackage{hyperref}

\hypersetup{
    colorlinks,
    citecolor=blue,
    filecolor=black,
    linkcolor=blue,
    urlcolor=blue
}

\usepackage{overpic}
\usepackage{array}
\usepackage{rotating}
\usepackage{datetime}

\newcommand{\bea}{\begin{eqnarray}}
\newcommand{\eea}{\end{eqnarray}}
\newcommand{\be}{\begin{equation}}
\newcommand{\ee}{\end{equation}}

\title{On Carroll partition functions and flat space holography}
 \author[a]{Georgios Poulias}
\author[a]{and Stefan Vandoren}

\affiliation[a]{Institute for Theoretical Physics, Utrecht University, \\ Princetonplein 5, 3584 CE Utrecht, The Netherlands}

\emailAdd{s.j.g.vandoren@uu.nl}

\abstract{We continue the study of Carroll limits on partition functions of relativistic conformal theories and their thermodynamics. By introducing imaginary chemical potentials $v$ conjugate to momenta, one can access and study the Carroll regime in which $v\gg c$. We analyze examples of free massless particles, 2d CFTs and some free field theories in $d$ spatial dimensions. For 2d holographic CFTs, we revisit and clarify further the connection between the Carroll limit and the flat space limit of BTZ black holes. The general picture, valid in any dimension, is that, starting from AdS/CFT, the boundary Carrollian field theory lives on the horizon of a large black hole in AdS, which is pushed to the null boundary in the flat space limit.

}

\begin{document}
                                           \maketitle

\section{Introduction}

The Carroll limit is commonly stated as the limit of vanishing speed of light, $c\to 0$ \cite{Levy1965,SenGupta:1966qer}. In such a limit, the lightcone closes up and any relativistic particle with velocity $v<c$ then freezes. The theory becomes ultralocal and remarkably this is relevant for flat space holography since the conformal Carroll group is isomorphic to the asymptotic BMS symmetry group in Minkowski space in one dimension higher \cite{Duval:2014uva}.  The Carroll algebra contains, next to the rotations, also translations in time (generated by $H$) and space $P_i$, and the Carroll boosts $C_i$, with commutation relations
\begin{equation}
    [P_i,C_j]=\delta_{ij}H\ ,\qquad [P_i,H]=[C_i,H]=0\ .
\end{equation}
In particular this means that the Hamiltonian is a central charge, since it commutes with all generators. This is different to relativistic theories where $H$ is not boost invariant and leads to the immediate puzzle that partition functions and density operators of the form
\begin{equation}
    Z={\rm Tr}\Big[e^{-\beta H}\Big]\ ,\qquad \rho_{\beta}=\frac{e^{-\beta H}}{Z}\ ,
\end{equation}
are not well defined as the trace typically diverges when $H$ is proportional to the unity operator. This problem was already signaled in \cite{deBoer:2023fnj}.
A follow up question is then how flat space holography works as, just like for AdS/CFT holography, the bulk and boundary partition functions are supposed to be the same.

One attempt to understand this, is to start from AdS/CFT, and consider the flat space limit in the bulk. On the boundary, this amounts to using $c$ as a regulator. For any small value of $c$ the boundary CFT partition function is well-defined, but divergences appear in the strict $c\to 0$ limit. As we will see, this has a physical interpretation when starting from AdS$_{D+1}$/CFT$_D$ holography, and then taking the large AdS radius limit $R\to \infty$ to flat space. Then, the partition function is also divergent in the $R\to \infty $ limit because of the presence of large black hole in AdS. That is because the black hole entropy scales like $r_+^{D-1}$, and the outer horizon tends to infinity in the flat space limit of large black holes (which, by definition, have $r_+\gg R\to \infty$). It is then natural to expect that the null surface on which Carroll field theories live, is actually the blown up event horizon of a large black hole in AdS. This is further supported by the fact that Carrollian symmetries naturally appear on black hole horizons \cite{Donnay:2019jiz} (see also \cite{Freidel:2024emv} for a more elaborate work).

But some care has to be taken when taking such a $c\to 0$ limit.  As usual, it is important to consider limits of dimensionless quantities, so one needs to identify one or more velocity scales $v\neq 0$ and consider the ratio $v/c$ to be very large in the Carroll limit. Sometimes, as in field theories, this leads to different possibilities for the Carroll limit, dubbed electric and magnetic Carroll limits \cite{Henneaux:2021yzg,deBoer:2021jej,Bergshoeff:2023vfd}, based on the analogy with the Carroll limit on electromagnetism \cite{Duval:2014uoa}.

In this paper, we study Carroll limits on thermodynamic systems, where new subtleties arise, as already pointed out in \cite{deBoer:2023fnj} and on which we elaborate. Since thermodynamic systems are at finite, large volume $V$ and temperature $\beta^{-1}=k_BT$, one can introduce a thermal velocity scale
\begin{equation}\label{v-th}
    v_{th}=\frac{L}{h\beta}=\frac{k_BT}{h}L\ ,
\end{equation}
where the volume in $d$ spatial dimensions is written as $V=L^d$, and $h$ is Planck's constant. Such a velocity is not bounded by $c$, and in the thermodynamic limit $v_{th}$ can be made arbitrary large\footnote{It is intriguingly similar to the Hubble-Lema\^itre law in cosmology, where the recessional velocity $v_{rec}=Hd$ can also exceed the speed of light beyond the cosmological horizon. For an application to Carrollian symmetry, see \cite{deBoer:2021jej}.}. Other velocities can be introduced by considering chemical potentials conjugate to momenta, by replacing the Hamiltonian with
\begin{equation}
H\to H-\vec v\cdot\vec P\ .
\end{equation}
As we will show, only after analytic continuation, $v\to i v$, can one consider Carroll limits of relativistic theories with $v\gg c$. The main aim of our paper is to show that this is a well-defined procedure by itself, and moreover it can be applied to flat space holography. In the context of AdS$_3$/CFT$_2$, we show in detail how the flat space limit corresponds to the Carroll limit as described above, and we connect to known results in flat space holography as already described in \cite{Barnich:2012aw,Barnich:2012xq,Bagchi:2012xr,Bagchi:2013lma}. 
 This is the subject of section 3, where we also discuss properties of the Carroll stress tensor and the first law of thermodynamics. In section 2, we treat simple quantum mechanical systems to illustrate the procedure of analytic continuation of $v$ and show the consistency of the Carroll limit in which $c/v\to 0$. And in section 4, we repeat this for free scalar field theories in higher dimensions compactified on a circle.

 More references on Carrollian physics connecting to flat space holography can be found e.g. in the review-like papers \cite{Donnay:2022aba,Donnay:2022wvx,Bagchi:2023cen} and references therein, and in some of the more recent references \cite{Campoleoni:2022wmf,Bagchi:2023fbj,Nguyen:2023vfz,Grumiller:2023rzn,Campoleoni:2023fug,Mason:2023mti,Have:2024dff,Bekaert:2024itn,Alday:2024yyj,Ruzziconi:2024zkr,Kraus:2024gso,Bagchi:2024gnn,Ruzziconi:2024kzo,Cotler:2024cia,Kulp:2024scx,Kraus:2025wgi,Nguyen:2025sqk}.

\section{Quantum mechanics}

In this section, we extend the analysis done in \cite{deBoer:2023fnj}, namely computing the partition function for massless particles in the Carroll regime. We generalize the computation of \cite{deBoer:2023fnj} to arbitrary dimensions and furthermore, we investigate in more detail the properties of pressure and energy density in the Carroll limit and how it leads to equations of state with $w=-1$. 

Part of our analysis holds more general for any quantum mechanical system with translation symmetry. In that case the Hamiltonian $H$ and momentum operator $\vec P$ commute, so we can define another conserved quantity
\begin{equation}
    H'=H-\vec v\cdot \vec P\ ,
\end{equation}
with $\vec v$ a generalized chemical potential conjugate to the momentum which has the dimensions of a velocity. A density matrix and partition function in the presence of such a chemical potential is
\begin{equation}
    Z={\rm Tr}\Big[e^{-\beta H+\beta \vec v\cdot \vec P}\Big]\ ,\qquad \rho_{\beta,v}=\frac{e^{-\beta H+\beta \vec v\cdot \vec P}}{Z}\ .
\end{equation}
One may call the density matrix at $v=0$ the thermal state in the rest frame, and otherwise in the lab frame. $H'$ and $H$ are related by a Galilei boost ${\vec x}^{\,'}=\vec x +\vec v\, t$.

In this paper, we will consider the analytic continuation $\vec v\to i\vec v$. For the partition function, this is quite natural since it can then be understood as one with the translation operator inserted,
\begin{equation}
    T_{\vec x}=e^{\frac{i}{\hbar}{\vec x}\cdot\vec P}=e^{i\beta{\vec v}\cdot\vec P}\ ,
\end{equation}
with the distance in the translation converted into the velocity 
\begin{equation}\label{v-x}
    \vec v=\frac{k_BT}{\hbar}\vec x\ .
\end{equation}
The partition function is then the Fourier transform of $e^{-\beta H(p)}$ and is important for understanding the spatial distribution of thermal states. When $P$ (say in one spatial dimension) has discrete eigenvalues, say $\hbar n/R$ if we are on a circle, then $x$ shows a discrete shift symmetry $x\to x+2\pi R$, corresponding to $v\to v+2\pi R/\hbar\beta=v+2\pi v_{th}$, with $v_{th}$ given by \eqref{v-th}. The angle $\varphi\equiv v/v_{th}$ has then period $2\pi$. Since we are on a circle, or on a torus in higher dimensions, the translation operator actually generates a rotation along the circle.

Having an imaginary chemical potential is standard in a Euclidean 2D CFT and important to show modular invariance of the partition function, as we review in the next section. A more controversial operation is to consider this analytic continuation on the density matrix $\rho_{\beta,v}\to \rho_{\beta,iv}$, since then $\rho$ is no longer hermitian. If we add the hermitian conjugate, then one obtains $\cos(\frac{{\vec x}\cdot{\vec P}}{\hbar})$, but this is no longer positive definite. It is therefore not clear that the expectation value of observables of the form $\langle {\cal O}\rangle ={\rm Tr}\Big[{\cal O}\cdot\rho_{\beta,iv}\Big]$ has a physical meaning, unless one continues back to real chemical potential.

Our reason for doing this is that after the analytic continuation, the speed $v$ obtained from \eqref{v-x} can then be taken larger than the speed of light $c$, and this is relevant for defining a Carroll limit. One of the purposes of this paper is to consider explicit examples where we can calculate everything, and to see what makes sense. 

In the remainder of this section, we study massless relativistic particles with single particle Hamiltonian given by
\begin{equation}
    H(\vec p)=c\,|\vec p|\ .
\end{equation}
In a basis of eigenstates of the momentum, the eigenvalues of $H'$ with real $v$ are given by
\begin{equation}
    E'=c|\vec p|-\vec v\cdot \vec p\ ,
\end{equation}
which is only positive for $v<c$. Such a generalized chemical potential can be generated by a Lorentz boost, since $H$ transforms under a boost with parameter $\vec v$, into
\begin{equation}
    H\to\gamma_v(H-\vec v\cdot \vec P)\ ,\qquad \gamma_v=\frac{1}{\sqrt{1-\frac{v^2}{c^2}}}\ .
    \end{equation}
Our aim is to compute the partition function, energy, momentum and pressure in a thermal ensemble, and then consider the Carroll limit. If we take the Carroll limit at the level of the Hamitonian, $H\to 0$ and we immediately see that, for real chemical potential, there is a divergence from integrating over momenta. Therefore we regulate it first by introducing a small speed of light, and at the end take a limit, loosely speaking $c\to 0$.

\subsection{Real chemical potential}

    We now consider the partition function for this system, defined as
\begin{equation}\label{Z1}
    Z_1(V,\beta,v^i)={\rm Tr}\Big[e^{-\beta H+\beta \vec v\cdot \vec P}\Big]=\frac{V}{h^3}\int {\rm d}^3\vec{p}\, e^{-\beta c|\vec{p}|+\beta v^ip_i}\ ,
\end{equation}
where in the second equation we regulated the system in the infrared by putting it in a large volume $V$ such that the momentum sum can be well approximated by an integral. 

A finite and fixed volume and temperature breaks Lorentz invariant, but one could formally make the partition function Lorentz invariant if we Lorentz contract the volume and use a boosted inverse temperature by writing it as $\beta=\gamma_v \beta_0$, where $\beta_0$ is the invariant inverse temperature. As in most literature, we will not do so, and in the following we take $\beta$ to be independent of $v$ \cite{deBoer:2023fnj}. 

For generic $v<c$, the integral in the partition function is convergent and we find 
    \begin{align}
        Z_1=\frac{8\pi V\gamma_-^4}{h^3\beta^3c^3}\ , \qquad \gamma_-\equiv \frac{1}{\sqrt{1-\frac{v^2}{c^2}}}\ .    
        \end{align}
This partition function can indeed formally be made Lorentz invariant by writing $\beta=\beta_0\gamma_-$ and the volume $V=V_0/\gamma_-$ with $V_0$ the proper volume.  If we do so, all $v$-dependence drops out and we are de facto in the rest-frame.

Because we are at finite volume and nonzero temperature, there are actually two kinds of velocities in the game. One is the chemical potential $v^i$, and one is the thermal velocity $v_{th}=L/h\beta$, as introduced in \eqref{v-th}\footnote{\label{largevth} This thermal velocity should not be confused with the thermal velocity associated to a massive particle, namely $v_{th}\approx {\sqrt{\frac{k_BT}{m}}}$, where $m$ is the mass of the particle. For massless particles, the thermal velocity defined by \eqref{v-th} is a velocity scale that can take any positive value and is not bounded by $c$. Plugging in the value of Planck and Boltzmann's constants, we have $v_{th}\approx 2\times 10^{10} (T/K) (L/s)$. e.g. for room temperature $T=300K$ and macroscopic size $L=1m$ we get much larger speeds then the speed of light, $v_{th}\approx 2\times 10^4\, c$.}. 
The partition function takes on the simple expression
\begin{align}\label{Z1minus}
        Z_1=8\pi \Big(\frac{v_{th}}{c}\Big)^3\gamma_-^4\ .
    \end{align}
While $v_{th}$ can be made arbitrary large, the velocity $v$ remains bounded to values $v<c$, otherwise the partition function is not convergent. In the Carroll limit, $v$ must therefore vanish. Then we have only one dimensionless ratio containing the speed of light, 
\begin{equation}
    y\equiv \frac{c}{v_{th}}=\frac{\beta}{L}hc\ ,
\end{equation}
and one can consider the Carroll limit $y\to 0$. For $v=0$, one has
\begin{equation}
    Z_1=\frac{8\pi}{y^3}\ ,
\end{equation}
and this is simply the same answer as for the relativistic case in the rest frame where $v=0$. It diverges in the strict Carroll limit $y\to 0$\footnote{It has been suggested in e.g. \cite{Ecker:2024czh} to rescale Planck constant $h\to h/c$, such that $y$ stays fixed in the Carroll limit. Even if one does so, the answer for the partition function remains that of a relativistic system, not that of a Carroll system.  Moreover, to avoid overall confusion, it is best to take limits on dimensionless quantities.}, so the most one can do is to keep $y$ small but non-zero which then is the same as the large temperature limit of the relativistic system.

We can compute expectation values of various quantities in this ensemble, by inserting operators in the trace of the density matrix. An example is the expectation value of the momentum operator. For a real chemical potential with $v<c$ we have
\begin{equation}\label{av-mom}
 \langle P_i \rangle_{1,v<c}\equiv\frac{1}{\beta}\frac{\partial}{\partial v^i}\ln Z_1=\frac{4}{\beta} \frac{\gamma_{-}^2}{c^2}v_i\ ,
 \end{equation}
and one could interpret this as the relativistic momentum of a (thermal) particle with mass that can be read of from the standard relation $\vec p=m\gamma_-\vec v$ with $mc^2=4k_BT$ kept fixed in the Carroll limit.

The partition function for $N$ indistinguishable particles can be well approximated, for low enough densities\footnote{Low enough densities compared to the thermal wavelength, $n\equiv N/V<<\lambda_{th}^{-d}$, where $\lambda_{th}\equiv hc\beta$ for massless particles. We can write this in terms of the thermal velocity $\lambda_{th}=(c/v_{th})L=yL$ such that low enough densities becomes the condition that $N<<y^{-d}$. In the Carroll limit where $y\to 0$, this is automatically satisfied.}, by 
\begin{equation}
    Z(N,T,V,v^i)=\frac{1}{N!}(Z_1)^N\ .
\end{equation}
The total average momentum in the canonical ensemble is then $N\langle P_i \rangle_1$, and the pressure $P_{pr}$ follows from
\begin{equation}
    P_{pr}=\frac{1}{\beta}\frac{\partial}{\partial V}\ln Z=\frac{N}{\beta V}\ ,
\end{equation}
which is the ideal gas law, and independent of $v$.
We now define the internal energy $\tilde E$ as
\begin{equation}\label{genE}
    \tilde E= -\frac{\partial \ln Z}{\partial \beta}=\frac{3N}{\beta}=3P_{pr}V\ . 
\end{equation}
The quantity $\tilde E$ is the expectation value of $H-\vec v\cdot \vec P$ and is independent of $v$. Hence it is the energy in the rest frame where $v=0$. The energy in the lab-frame where $v\neq 0$ is the expectation value of $H$ and
is computed from
\begin{equation}\label{Ev<c}
    E=\tilde E +\vec v\cdot\vec P=\frac{3N}{\beta}+\frac{4N}{\beta}\frac{v^2}{c^2}\gamma_-^2\ ,
\end{equation}
which is manifestly positive. In the Carroll limit, in which $v^2<c^2\to 0$, the lab-frame and rest frame results are the same and one finds the usual equation of state for an ideal gas of conformal matter in $d$ spatial dimensions, ${\cal E}=d P_{pr}=dNk_BT$, where ${\cal E}=E/V$ is the energy density. This thermal energy is linear in the temperature, is independent of $c$ and so it survives in the Carroll limit. For zero temperature, the energy density vanishes.

Curiously, the expression for the partition function \eqref{Z1minus} can be continued to values $v>c$ and is well-defined it this regime. This regime however cannot be obtained from the partition function with real chemical potential. We now show how to resolve this.

\subsection{Imaginary chemical potential}
    
For chemical potentials $v>c$, the integral over momenta in \eqref{Z1} does not converge. One way to make it convergent is  to make $v$ imaginary \cite{deBoer:2023fnj} and define the partition function\footnote{In all of this section, we assume here $v\neq 0$. The case $v=0$ is treated in the previous subsection, and leads to divergences in the strict Carroll limit.} 
\begin{equation}
     Z_1=\frac{V}{h^3}\int {\rm d}^3{\vec p}\,e^{-\beta c|\vec{p}|+i\beta v^ip_i}= 8\pi \Big(\frac{v_{th}}{c}\Big)^3\gamma_+^4\ ,\qquad \gamma_+\equiv \frac{1}{\sqrt{1+\big(\frac{v}{c}\big)^2}}\ ,
\end{equation}
which is the analytic continuation of the relativistic answer with real chemical potential when replacing $v\to iv$. After the analytic continuation, the partition function is still real but now defined for any value of $v$. Since it is no longer a boost parameter, $v$ is not bounded by $v<c$. 

For imaginary chemical potential, we compute
\begin{equation}\label{av-mom+}
    \langle P_i \rangle_{1}\equiv -i\frac{1}{\beta}\frac{\partial}{\partial v^i}\ln Z_1=i\frac{4}{\beta} \frac{v_i}{c^2}\gamma_{+}^2\ ,
\end{equation}
which is the analytic continuation $v\to iv$ of \eqref{av-mom}, as it should, but now also valid for $v>c$. As a consequence of our definition in \eqref{av-mom+},  the average momentum is now purely imaginary. The operator $\vec P$ is still hermitian, but we get imaginary expectation value because the state in which we compute its expectation value is based on the density matrix with an imaginary chemical potential. The thermal mass then picks up a factor of $i$ and one could use the terminology that the thermal photon becomes tachyonic because its thermal mass squared becomes negative.

We now consider the Carroll limit on the partition function. For $v\neq 0$, the partition function contains two dimensionless parameters containing the speed of light, namely
\begin{equation}
    x\equiv \frac{c}{v}\ , \qquad y\equiv \frac{c}{v_{th}}\ .
\end{equation}
We can then take the Carroll limit $x\to 0$ and $y\to 0$ keeping 
\begin{equation}
    \frac{x}{y}=\frac{v_{th}}{v}=\frac{L}{h\beta v}\ ,
\end{equation}
fixed. Small $x$ requires an imaginary chemical potential because $v>c$. The leading order expansion of the partition function then yields
\begin{equation}
    Z_1 = 8\pi \Big(\frac{x}{y}\Big)^3\, x\Big(1-4x +{\cal O}(x^2)\Big)\ .
\end{equation}
This vanishes in the strict Carroll limit\footnote{If we take the Carroll limit in the exponent of the partition function, the Hamiltonian vanishes and the momentum integral yields a delta function $\delta(\vec v)$, which for $v\neq 0$ vanishes.}, so we keep the leading non-zero term in the small $x$ expansion, which is called the Carroll regime. For the average momentum with $v\neq 0$, the strict Carroll limit is actually non-zero and finite,
\begin{equation}
    \langle P_i \rangle_1 \to i\frac{4}{\beta v} \frac{v_i}{v}=\frac{4i\hbar}{R}\frac{v_{th}}{v}n_i\ ,
\end{equation}
where $n_i=v_i/v$ is a unit vector. Note that in the limit $v\to \infty$ ($v\gg v_{th}$), the average momentum vanishes.

Before we continue with the energy density and pressure, we first generalize to arbitrary dimensions, as was done in \cite{Poulias}. The result for the partition function is again a Fourier transform and the result is
\begin{equation}
    Z_1(V,T,v)=
    \frac{2^d\pi^{\frac{d-1}{2}}V}{(h\beta c)^d}\gamma_{\mp}^{d+1}\Gamma\Big(\frac{d+1}{2}\Big)\ ,
\end{equation}
where $\Gamma$ is the Gamma function, and which holds for both cases (real chemical potential with $v<c$, or imaginary chemical potential with $v$ arbitrary). In the latter case, we have to choose $\gamma_+$, and in the Carroll limit, we get as a leading term (with $V=L^d$)
\begin{equation}
    Z_1 \to 2^d\pi^{\frac{d-1}{2}} \Gamma\Big(\frac{d+1}{2}\Big)   \Big(\frac{x}{y}\Big)^d x\ .
    \end{equation}
As before, this vanishes in the strict Carroll limit $x\to 0$ with $x/y=L/h\beta v$ fixed.

\begin{figure}
    \centering
    \includegraphics[width=0.8\linewidth]{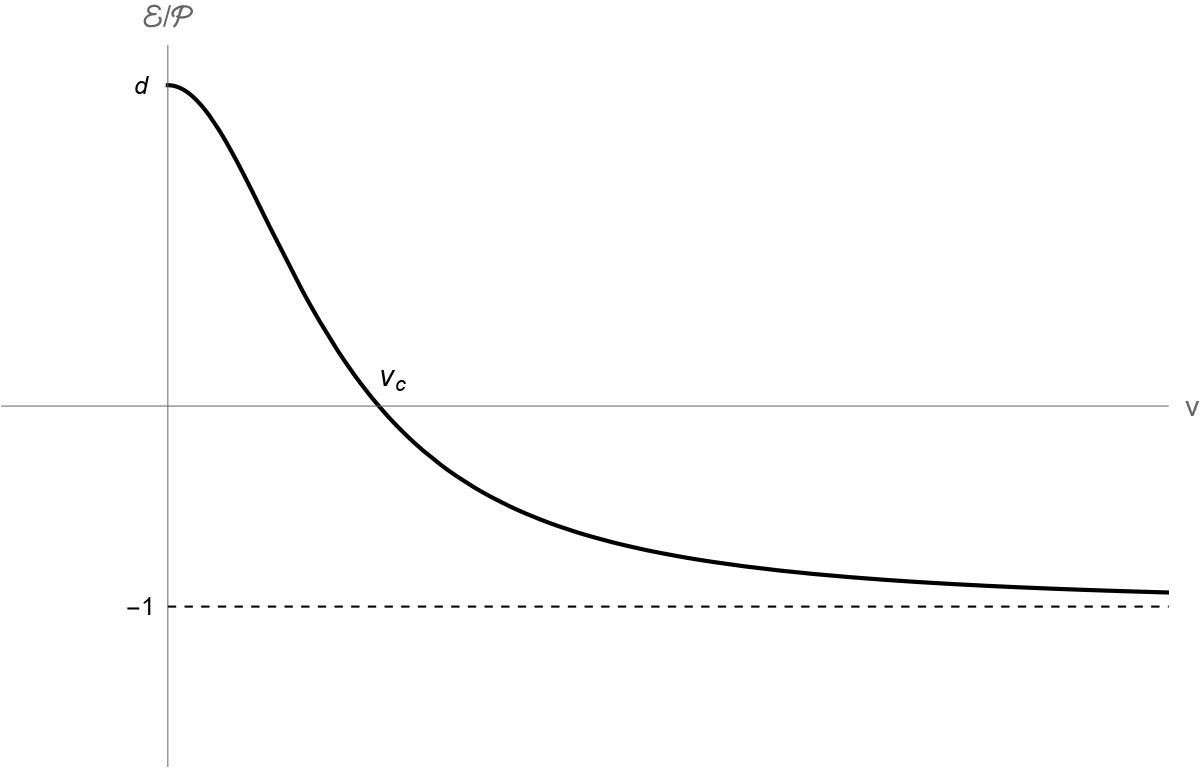}
    \caption{Energy density over pressure vs the generalized chemical potential}
    \label{fig:figureEOS}
\end{figure}

For the partition function of $N$ particles and the case of imaginary chemical potential, the pressure remains $P_{pr}=N/\beta V$ and for the average energy we have
\begin{equation}\label{C-Energy-particle}
    E=\tilde E +i\vec v\cdot\vec P=\frac{dN}{\beta}-\frac{(d+1)N}{\beta}\frac{v^2}{c^2}\gamma_+^2\ .
    \end{equation}
    This is again the analytic continuation of \eqref{Ev<c} but as a consequence, it is no longer positive for all values of $v$. It is finite for all values of $v$, e.g. for $v=c$, we get $E=(d-1)N/2\beta$. In the Carroll limit, where $\gamma_+^2\to c^2/v^2$, the energy stays finite and becomes
    \begin{equation}
        E \to E_{{\rm Carroll}}=-\frac{N}{\beta}=-P_{pr}V\ ,
    \end{equation}  
    which is negative.
        In terms of the pressure and energy density ${\cal E}=E/V$ and $x=c/v$, we can write \eqref{C-Energy-particle} as
    \begin{equation}\label{EOS}
        \frac{{\cal E}}{P_{pr}}=d-\frac{d+1}{1+x^2}=\frac{dc^2-v^2}{c^2+v^2}\hspace{0.2cm}.
\end{equation}
This is illustrated by Figure~\ref{fig:figureEOS}.
Something remarkable happens in the Carroll limit $x\to 0$ as we find, for any dimension, an equation of state $P_{pr}=w{\cal E}$ with $w=-1$. Such an equation of state is not possible with ordinary matter because both energy and pressure are positive. For Carrollian matter, the energy density can be negative (in our case, $E=-N/\beta$), and so there is the possibility to have $w=-1$, as also follows from the general analysis in \cite{deBoer:2017ing}. As is easily checked, the energy density becomes negative for values $v>v_c={\sqrt d} c$. For generic values of $x$ one gets an equation of state with
    \begin{equation}\label{EOS}
        w=-\frac{1+x^2}{1-dx^2}=-1-(d+1)x^2+\cdots\ .
    \end{equation}
    Notice furthermore that for large $x=c/v$, one gets back the relativistic answer for conformal matter, namely $w=1/d$. This corresponds to the case of $v=0$, as mentioned above. At the point $v_c=\sqrt{d}c$, $w$ tends to infinity, but this is simply signaling the fact that the energy density vanishes as follows easily from \eqref{C-Energy-particle}, and becomes negative for smaller values of $x$. The pressure stays finite and positive all the way, and is equal to $N/\beta V$ independent of $x=c/v$.

\section{Carrollian 2d CFT}

In this section, we present a similar analysis as for massless particles, but now in the context of two-dimensional conformal field theory (2d CFT). The main aspects of our analysis, namely Carroll limits on the low and high temperature behavior of 2d CFTs,  are universal and will hold for any 2d CFT and therefore apply as well to AdS$_3$/CFT$_2$ holography. For the non-universal part, and for illustrational purposes, we will consider a simple model, namely a two-dimensional CFT for $N$ massless real scalar fields, each with Lagrangian density 
\begin{equation}
    \mathcal{L}=\frac{1}{2c^2}({\partial_t\phi})^2-\frac{1}{2}(\partial_x\phi)^2=\frac{c^2}{2}\pi_\phi^2-\frac{1}{2}(\partial_x\phi)^2\ ,
\end{equation}
where $\pi_\phi=\frac{1}{c^2}\partial_t\phi$ is the conjugate momentum. The Carroll limit of a free scalar and its partition function was already studied in \cite{Hao:2021urq,deBoer:2023fnj} but here we develop further the thermodynamic properties and we make the connection to flat space holography by considering universal aspects of 2d CFT's at large central charge. In the language of representations, the obtained results for the Carroll limit of partition functions should be understood in terms of unitary induced representations, not the non-unitary highest weight representations of the BMS$_3$ algebra. See e.g. \cite{Bagchi:2019unf,Hao:2021urq} for more on the highest weight representations. 

Finally a disclaimer: many of the things that follow in the remaining of this section are known results. In section 3.1, we rewrite known results about 2d CFTs and their holographic duals, but in a language that involves the chemical potential $v$. In section 3.2, apart from the results on the stress tensor and the first law, everything about the flat space limit of the BTZ black holes is known from \cite{Barnich:2012aw,Barnich:2012xq,Bagchi:2012xr,Bagchi:2013lma}. Our aim here is to see in detail how the flat space limit matches with the Carroll limit $v\gg c$ on the CFT$_2$ that we introduced in this paper. This builds up confidence that our procedure of taking the Carroll limit is well-defined and can be used in higher dimensions as well.

\subsection{Preliminaries}

Before discussing Carroll limits, we review some general known facts about 2d CFTs, mostly for notational purposes and making factors of $c$ explicit. We put the theory on a circle of radius $R$ with $x\sim x+2\pi R$. After a Wick rotation to Euclidean space with $\tau=it$, we 
define complex coordinates 
\begin{equation}
    z=e^{\big(\frac{c}{R}\tau+i\frac{x}{R}\big)}\ ,
\end{equation}
such that
\begin{equation}
    \partial_\tau=\frac{c}{R}(z\partial_z+\bar z \partial_{\bar z})\ ,\qquad \partial_x=\frac{i}{R}(z\partial_z-\bar z \partial_{\bar z})\ .
\end{equation}
This leads to the operator identities (with $H=i\hbar\partial_t=-\hbar\partial_\tau$ and $P=-i\hbar\partial_x$)
\begin{equation}\label{H-P}
    H=\frac{c\hbar}{R}(L_0+\bar L_0)\ ,\qquad P=-\frac{\hbar}{R}(L_0-\bar L_0)\ ,
\end{equation}
with $L_0=-z\partial_z$ and $\bar L_0=-\bar z\partial_{\bar z}$. Notice the speed of light appearing in $H$ but not in $P$. The partition function with imaginary chemical potential\footnote{For later convenience we changed $v\to -v$ in the partition function \eqref{part-CFT} compared to the previous section. Changing $v\to -v$ only has the consequence of sending $\tau_1\to -\tau_1$ in the modular parameter, but $\tau_2$ remains positive.} can now be written as
\begin{equation}\label{part-CFT}
    Z(\tau,\bar\tau)={\rm Tr}\Big(e^{-\beta H-i\beta vP}\Big)={\rm Tr}\Big(q^{L_0-\frac{\rm c}{24}}\bar{q}^{\bar{L}_0-\frac{\bar{\rm c}}{24}}\Big)\ ,
\end{equation}
with
\begin{equation}\label{taumodpar}
  \tau\equiv\tau_1+i\tau_2=\frac{\beta v\hbar}{2\pi R}+i\frac{\beta c \hbar}{2\pi  R}\ ,  \qquad q=e^{2\pi i\tau}\ .
\end{equation}
We took into account the Casimir energy of the CFT with, for convenience, equal left and right central charges ${\rm c}=\bar{\rm c}$\footnote{We use the somewhat perverse notation to denote the speed of light with $c$ and the central charge with a roman c. Furthermore, c denotes the left moving central charge, so for the free scalar we have c=1.}, 
\begin{equation}\label{casimir}
    E_C=-\frac{c\hbar}{12R}{\rm c}=-\frac{\pi{\rm c}}{6\beta}\tau_2\hspace{0.2cm}.
\end{equation}
Hence we see that having imaginary chemical potentials is standard in two-dimensional CFTs, as they correspond to rotations along the circle. See e.g. also \cite{Alessio:2021krn} for a recent reference related to our discussion. The Euclidean CFT at finite temperature is now defined on a torus with modular parameter $\tau$ given by \eqref{taumodpar} which holds for any CFT$_2$. Similarly the formula for the Casimir energy \eqref{casimir} holds for any CFT$_2$. We can therefore consider the case relevant for AdS$_3$/CFT$_2$ holography\footnote{Our conventions are as follows. The metric on AdS$_3$ in global coordinates is ${\rm d}s^2=-(1+\frac{r^2}{R^2})c^2{\rm d}t^2 +(1+\frac{r^2}{R^2})^{-1}{\rm d}r^2+r^2{\rm d}\varphi^2$, with $\varphi \sim \varphi+2\pi$. Close to the boundary $r\gg R$, the metric is conformal to a cylinder with metric ${\rm d}s^2=-c^2{\rm d}t^2+R^2{\rm d}\varphi^2$. Hence the boundary field theory lives on a spatial circle with radius $R$.} if we start with the Brown-Henneaux formula for the central charge \cite{Brown:1986nw} 
  \begin{equation}\label{BrownHenneaux}
      {\rm c}=\frac{3R}{2G}\ ,
  \end{equation}
  with $G$ the 3d Newton's constant with units of length, and $R$ the AdS$_3$ radius. The Casimir then becomes the AdS$_3$ energy \cite{Strominger:1997eq},
  \begin{equation}
      E_C=-\frac{c\hbar}{8G}=E_{AdS_3}\ .
  \end{equation}
Corrections to the zero temperature limit in the CFT depend on the model. We can compute them explicitly for ${\rm c}=N$ free bosons.  The partition function is determined by the modular (of weight 1/2) Dedekind eta form, 
\begin{equation}\label{Dedekind}
    Z(\tau,\bar\tau)=\frac{1}{(\sqrt{\tau_2})^N}\frac{1}{|\eta(q)|^{2N}}\hspace{0.2cm}.
\end{equation}
It is easy to see that the low temperature regime again becomes independent of the chemical potential $v$, as we have for small $q$ that $\eta(q)\sim q^{1/24}$, and so
\begin{equation}\label{DedekindlowT}
    Z(\tau,\bar\tau)=\frac{1}{(\sqrt{\tau_2})^N}e^{-\beta E_C}\ ,
\end{equation}
without any dependence on $\tau_1$ (and hence $v$).
Hence for small temperatures, we find for energy density ${\cal E}=E/R$ and pressure
\begin{equation}
{\cal E}=P_{pr}=\frac{1}{R}\Big(-E_C+\frac{N}{2}k_BT\Big)\ ,
\end{equation}
corresponding to an equation of state with $w=1$ and consistent with conformal matter in one spatial dimension. The zero temperature limit is universal and states the well-known result that pressure is minus the Casimir energy density. 

The large temperature limit brings us into the Cardy regime. This corresponds to $\tau\to 0$ (and hence $q\to 1$), and the result shows dependence on $v$. We can find the behavior for the Dedekind eta function around $\tau \approx 0$ from the identity $\eta(-1/\tau)={\sqrt{-i\tau}}\eta(\tau)$, together the the well-known $q$-expansion $\eta(\tau)=q^{\frac{1}{24}}(1-q-q^2+\cdots )$. The small $q$ expansion corresponds to $\tau_2\to \infty$, but thanks to the modular properties we can evaluate $\eta(\tau)$ around $\tau\to 0$, 
\begin{equation}
    \eta(\tau)\to \frac{e^{-\pi i/12 \tau}}{{\sqrt{-i\tau}}} \ ,
\end{equation}
and from this we can find the  high-$T$ limit of the partition function for $N$ free scalars,
\begin{equation}
    Z=\frac{1}{(\sqrt{\tau_2})^N}\frac{1}{|\eta(\tau)|^{2N}}\overset{\tau\rightarrow 0}{\rightarrow}\Big(\frac{|\tau|}{\sqrt{\tau_2}}\Big)^N e^{\frac{\pi i}{12}(\frac{1}{\tau}-\frac{1}{\bar\tau})N}\ .
\end{equation}
The logarithm of the partition function in the large temperature limit is (switching again to the central charge c$=N$)
\begin{equation}\label{lnZCardy}
    \ln Z\to {\rm c}\ln\frac{|\tau|}{\sqrt{\tau_2}}+\frac{\pi \tau_2}{6|\tau|^2}{\rm c}\ .
\end{equation}
Since $\tau\sim 1/R$, the second term is linear in $R$ and shows extensivity. For sufficiently large $R$ this is the dominant term,
\begin{equation}\label{ZCardy}
    \ln Z\to \frac{{\rm c}}{3}\frac{\pi^2 R}{\beta\hbar c}\frac{1}{1+v^2/c^2}\ , \qquad \frac{\tau_1}{\tau_2}=\frac{v}{c}\ .
\end{equation}
This is again a universal result, that follows from the Casimir energy at low temperature by using modularity $\tau\to -1/\tau$. Using the Brown-Henneaux formula \eqref{BrownHenneaux}, the corresponding entropy is
\begin{equation}\label{CardyEntropy}
    \frac{S}{k_B}=\frac{\pi^2 R^2}{G}\frac{k_BT}{\hbar c}\frac{1}{1+\frac{v^2}{c^2}}=\frac{\pi^2 R}{G}\frac{v_{th}}{c}\frac{1}{1+\frac{v^2}{c^2}}    
    \ .
\end{equation}
We can now match this with the entropy of the rotating BTZ black hole \cite{Banados:1992wn,Banados:1992gq}, which can be written as 
\begin{equation}
    \frac{S_{BTZ}}{k_B}=\frac{\pi r_+}{2 G}=\frac{\pi^2R^2}{G}\frac{k_BT}{\hbar c}\frac{1}{1+\frac{|r_-|^2}{r^2_+}}\ ,
\end{equation}
where the inner and outer horizons of the black hole are denoted by $r_\pm$ and with  Hawking temperature given by,
\begin{equation}
    \frac{k_BT}{\hbar c}=\frac{r_+^2-r_-^2}{2\pi R^2 r_+}\ .
\end{equation}
In terms of the mass and angular momentum, the horizon formula read (we follow the conventions of \cite{Barnich:2012aw})
\begin{equation}\label{horizons}
    r_\pm^2=4GMR^2\Big(1\pm\sqrt{1+\frac{J_E^2}{M^2R^2}}\Big)\ ,
\end{equation}
 with $M$ having units of inverse length, such that $\hbar c M$ has units of energy, and $J$ is dimensionless. Notice that 
we have used the Euclidean BTZ solution (the 2d CFT is also Euclidean) with Wick rotated angular momentum $J_L=iJ_E$ such that $r_-^2<0$ for real $J_E$. Hence we defined, as is well-known in the literature (see e.g. \cite{Carlip:1994gc}), $|r_-|=ir_-$. The matching with the CFT is then complete if we identify, not unexpectedly, the velocity $v$ with the angular velocity of the outer black hole horizon ($\Omega=c|r_-|/r_+$),
\begin{equation}\label{v/c-holo}
    \frac{\tau_1^2}{\tau_2^2}=\frac{v^2}{c^2}=\frac{|r_-|^2}{r_+^2}\leq 1\ .
\end{equation}
The case of $v=0$ then corresponds to the non-rotating BTZ black hole with $r_-=0$.

We can also compute the pressure
\begin{equation}\label{pressure}
    P_{pr}=\frac{1}{\beta}\frac{\partial \ln Z}{\partial R}=\frac{\pi^2}{3\hbar\beta^2}\frac{c}{v^2+c^2}{\rm c}\ ,
\end{equation}
and average momentum 
\begin{equation}
    P_{mom}=i\frac{1}{\beta}\frac{\partial}{\partial v}\ln Z=-\frac{2i}{3}\frac{\pi^2R}{\beta^2\hbar}\frac{cv}{(c^2+v^2)^2}{\rm c}\ .
\end{equation}
The generalized energy is then
\begin{equation}
    \tilde E= -\frac{\partial \ln Z}{\partial \beta}=\frac{\pi^2R}{3\hbar\beta^2}\frac{c}{v^2+c^2}{\rm c}=P_{pr}R\ .
\end{equation}
The relation $\tilde {\cal E}=P_{pr}$ is a consequence of the vanishing of the trace of the stress tensor of any 2d CFT. For $v=0$, $\tilde E$ does become the energy, and we get the usual CFT relation for the energy density ${\cal E}=P_{pr}$. For $v\neq 0$, the quantity $\tilde E$ is not the actual energy, but, just as for the particle case, we have (taking into account the change in notation $v\to -v$), $E={\tilde E}-ivP_{mom}$, and with energy density ${\cal E}=E/R$, we find
\begin{equation}\label{EPrelation2dCFT}
    {\cal E}=\frac{c^2-v^2}{c^2+v^2}P_{pr}\ ,
\end{equation}
and we remind that this is the result for the relativistic CFT with non-zero, imaginary chemical potential $v$. Interestingly, the equation of state parameter $w$ does not depend on the central charge, radius and temperature. In the absence of chemical potential, $v=0$, we have again $w=+1$ and the energy density simplifies 
\begin{equation}
    {\cal E}=\frac{\pi^2{\rm c}}{3\hbar c}\,(k_BT)^2\ .
\end{equation}
All of the above is known, and we just rewrote some known formulars in terms of the chemical potential $v$. This is to prepare for the Carroll limit to which we turn now.

\subsection{Carroll limit}

  We now discuss the Carroll limit and focus mainly on the universal aspects at low and high temperature. Using the formula for the thermal velocity, we can write the modular parameter as
\begin{equation}
  \tau=\frac{v}{2\pi v_{th}}+i\frac{c}{2\pi v_{th}}\ ,\qquad  v_{th}=\frac{R}{\beta \hbar}\ ,
\end{equation}
so we have the usual relation that $\varphi\equiv v/v_{th}=2\pi {\rm Re}\tau\in [0,2\pi]$ up to $\tau\to \tau+1$. The Carroll limit involves sending $\tau_2\to 0$, so $q\to e^{2\pi i\tau_1}$. The strict Carroll limit $\tau_2=0$ is not well-defined because $\tau$ is defined on the upper half plane where $\tau_2>0$, so it only makes sense to talk about the Carroll regime in which $\tau_2$ is small but non-zero. It is then easy to see what happens to modular transformations in the Carroll limit. The T-generator is still as above, $\tau_1\to \tau_1+1$, and the S-generator $\tau\to -1/\tau$ becomes in the Carroll limit (here we assume $\tau_1\neq 0$)
\begin{equation}\label{CarrollModular}
    \tau_1\to -\frac{1}{\tau_1}\ ,\qquad \tau_2 \to \frac{\tau_2}{\tau_1^2}\ .
\end{equation}
We expect Carroll partition functions to be modular with respect to this transformation. This Carrollian modular property was already put forward in \cite{Barnich:2012xq,Bagchi:2012xr} and demonstrated explicitly for the free scalar field in \cite{Hao:2021urq}. \\

{\textbf{Low temperature}}\\

At low temperature, the 2d CFT partition function is governed by the Casimir energy
\begin{equation}\label{casimir}
    E_C=-\frac{c\hbar}{12R}{\rm c}=-\frac{\pi{\rm c}}{6\beta}\tau_2\hspace{0.2cm}.
\end{equation}
For fixed central charge, the Casimir energy vanishes in the $\tau_2\to 0$ limit. However, in the context of holography, we use the Brown-Henneaux formula for the central charge \eqref{BrownHenneaux} and now propose a holographic Carroll limit by taking $\tau_2\to 0$, c$\to \infty$ such that c$\tau_2$ is fixed. This means $v_{th}\gg c$ in such a way that the Casimir energy $E_C$ stays constant, and one can now interpret the subscript $C$ for Carroll instead of Casimir. The partition function then is dominated by
\begin{equation}\label{Zflat}
   Z\propto
   e^{-\beta E_C}=e^{\frac{c\hbar\beta}{8G} }\ .
\end{equation}
Since Newton's coupling arises, this formula is now to be interpreted in the bulk.  Upon taking take the large radius, flat space limit $R\to \infty$, the Casimir remains finite (it simply does not depend on $R$), and therefore it remains the dominant contribution to the partition function in the Carroll limit. The result \eqref{Zflat} is actually known to be the leading (tree-level) term in the partition function of 3d gravity in thermal flat space \cite{Bagchi:2013lma} (see also \cite{Barnich:2015mui,Oblak:2015sea} for subsequent work on the partition function, and more recently \cite{Cotler:2024cia}), and therefore the Carroll limit on the partition function of the 2d-CFT is consistent with flat space holography. \\

\textbf{High temperature}\\

There are two cases to consider, the case of $v=0$ and $v\neq 0$. In both cases there is a Carroll limit. We first look at $v=0$ which corresponds to the BTZ black hole without angular momentum. It still has entropy given by $S\propto r_+\propto R$ which matches with the boundary CFT upon identifying $T$ with the Hawking temperature. The bulk entropy diverges in the flat space limit since $r_+^2=8MGR^2\to \infty$. Also on the boundary side, the entropy diverges, as  \eqref{CardyEntropy} gives
\begin{equation}
   v=0:\qquad \frac{S}{k_B}=\frac{\pi^2R}{G}\frac{v_{th}}{c}\ ,
\end{equation}
which diverges too in the Carroll limit $v_{th}\gg c$. This is going to be a generic feature of the Carroll/flat space limit. Large black holes in AdS have outer horizons $r_+\to \infty$, and the boundary Carroll field theory lives precisely on this null surface.

The case $v\neq 0$ is more interesting.  Now we start from the partition function \eqref{ZCardy} and consider the limit 
\begin{equation}
    v\gg c\ .
\end{equation}
The leading term in the partition function then becomes
\begin{equation}\label{CarrollZCardy}
    \ln Z\to \frac{{\rm c}\tau_2}{6}\frac{\pi}{\tau_1^2}\ ,
    \end{equation}
    which stays finite and non-zero in the Carroll limit. The entropy then follows easily,
\begin{equation}
    \frac{S}{k_B}=\frac{{\rm c}\tau_2}{3}\frac{\pi}{\tau_1^2}=
    \frac{\pi^2 R^2}{G}\frac{k_BT}{\hbar c }\frac{c^2}{v^2}\ ,
\end{equation}
where in the last equation, we have used the Brown-Henneaux formula for the central charge. If flat space holography applies, this formula should be matched with the entropy of a gravitational solution in asymptotically Minkowski space. The answer was given in \cite{Barnich:2012aw,Barnich:2012xq,Bagchi:2012xr,Bagchi:2013lma}, where it was shown that the flat limit of a rotating BTZ black hole becomes a cosmological solution in asymptotically Minkowski spacetime with metric in BMS coordinates
\begin{equation}
    {\rm d}s^2=8GMc^2{\rm d}u^2-2c{\rm d}u{\rm d}r+8GJ{\rm d}u{\rm d}\varphi+r^2{\rm d}\varphi^2\ ,
\end{equation}
and with the inner horizon of the BTZ black hole becoming the cosmological horizon. For $R\to \infty$, 
fixing $M$ and $J$, one finds from \eqref{horizons} that inner and outer horizons behave like
\begin{equation}
  r_+^2\to   8MGR^2\to \infty\ ,\qquad r_-^2\to -2G\frac{J_E^2}{M}\ .  
\end{equation}
The outer horizon still is blown up to the boundary where the Carroll field theory lives. The entropy should therefore not be matched with the flat limit of the outer horizon of the BTZ black hole (which required $v<c$!), but rather with the flat limit of the inner horizon
\begin{equation}
    \frac{S_{grav}}{k_B}=\frac{\pi |r_-|}{2 G}=\frac{\pi J_E}{\sqrt{2MG}}\ .
\end{equation}
The temperature associated with the inner horizon is obtained by interchanging $r_\pm$ in outer horizon expression,
\begin{equation}
    \frac{k_BT}{\hbar c}=\frac{r_+^2-r_-^2}{2\pi R^2 |r_-|}= \frac{{\sqrt{8G}}M^{3/2}}{\pi J_E}\ ,
\end{equation}
where in the last equation we took again the flat space limit.
Matching the bulk entropy with the Carroll dual, one finds the expression for the chemical potential
\begin{equation}\label{v/c}
    \frac{\tau_1}{\tau_2}=\frac{v}{c}=\frac{2M}{J_E}R\ ,
\end{equation}
which is also the expression for the angular velocity of the inner horizon in the flat space limit, 
\begin{equation}
v=\Omega_-=c\frac{r_+}{|r_-|}\to c\frac{2M}{J_E}R\ .
\end{equation}
From this identification, one sees again that the flat space limit in the bulk relates to the $c\to 0$ limit on the boundary. Observe furthermore that $v>c$ and this connects to the fact that the angular velocity of the inner horizon of the black hole is larger than the speed of light. The inner horizon of a black hole is known to suffer from instabilities. In the flat space limit this inner horizon becomes a cosmological horizon in a spacetime without cosmological constant, and there is nothing unstable about it (but see however the first law at the end of this section). 

Using \eqref{v/c}, the leading term in the Carroll partition function becomes
\begin{equation}
    \ln Z= \frac{\pi}{8G}\frac{J_E}{\sqrt{M}}=\frac{\pi}{4G}|r_-|\ ,
\end{equation}
matching the known result computed from the bulk in \cite{Bagchi:2013lma}.

Essentially most of this is a rewrite of the results obtained in \cite{Barnich:2012aw,Barnich:2012xq,Bagchi:2012xr,Bagchi:2013lma}, where more information can be found, e.g. about the phase transition between low and high temperatures \cite{Bagchi:2013lma}. What we add here are the details of the Carroll limit, and the consistency and interpretation of the partition function with imaginary chemical potential with $v\gg c$. A dual Carroll field theory analysis for the partition function was also considered in \cite{Bagchi:2012xr}, but that was based on highest weight, non-unitary representations. Here we have shown that the induced (and unitary) representations as obtained from a Carroll limit of a relativistic CFT, do the job and reproduces the correct entropy in the bulk. \\

\textbf{Carroll stress tensor}\\

The pressure of the CFT in the Carroll limit can be obtained from 

\begin{equation}\label{Carrollpressure}
    P_{pr}=\frac{1}{\beta}\frac{\partial \ln Z}{\partial R}=\frac{\pi^2}{3\hbar\beta^2}\frac{c}{v^2}{\rm c}\ .
\end{equation}
Since we keep $c\times {\rm c}$ fixed and finite, the Carroll limit is finite and non-zero. The pressure in the CFT has units of energy density, namely $J/m$. Using our dictionary, in terms of bulk data, we find
\begin{equation}
    P_{pr}=\frac{M\hbar c}{R}\ .
\end{equation}
Similary, one can compute the momentum in the Carroll limit and using the holographic dictionary we find
\begin{equation}
    P_{mom}=-\frac{\hbar J_L}{R}\ ,
\end{equation}
where we continued back to Lorentzian space with $J_L=iJ_E$. We then find the relation that $vP_{mom}/R=-2P_{pr}$.

The energy density of the 2d CFT was computed in \eqref{EPrelation2dCFT}. The equation of state parameter $w$ defined from $P_{pr}=w{\cal E}$ is follows as a function of the velocity $v$ 
\begin{equation}
    w=\frac{c^2+v^2}{c^2-v^2}=-\frac{1+\big(\frac{c}{v}\big)^2}{1-\big(\frac{c}{v}\big)^2}\hspace{0.2cm}.
\end{equation}
This relation is actually exactly the same as for massless particles with $d=1$, see \eqref{EOS}. As we showed in the previous section, for small values of the velocity $v$ compared to the speed of light, the equation of state parameter is $w=+1$ as expected ($w=\frac{1}{d}$) and in the Carroll limit $w=-1$. To first order in the small $c/v$ expansion, we find 
\begin{equation}
    w=-1-2\frac{c^2}{v^2}+\cdots\ .
\end{equation}
The flipping point where the energy changes sign is at $v=c$, so for all $v>c$ we have negative energy. This follows the general answer that for $v>{\sqrt d}c$, the energy becomes negative. 

Putting everything together, we can build up the Carroll stress tensor, which reads
\begin{equation}\label{CarrollstressCFT}
    T^\mu{}_\nu=\Big(\begin{matrix}
        P & 0 \\ -\frac{2P}{v} & -P
    \end{matrix}\Big)=\frac{1}{R}\Big(\begin{matrix}
        M\hbar c & 0 \\ -J_L\hbar & -M\hbar c
    \end{matrix}\Big)\ ,
\end{equation}
where $P=P_{pr}$ denotes the pressure and we used that ${\cal E}=-P$, and in the second equation we rewrote it in terms of bulk quantities. This matches  perfectly with flat space holography again as $M$ and $J$ are the $g_{uu}$ and $g_{u\phi}$ components of metric of the cosmological solution in BMS coordinates and which appear in the boundary BMS charges for Hamiltonian and momentum, as shown in \cite{Barnich:2012xq}. 

The stress tensor \eqref{CarrollstressCFT} can also be compared with the general form of the Carroll stress tensor obtained in \cite{deBoer:2017ing}, see eqns (2.58-61) in that reference, which in two dimensions reads
\begin{equation}\label{Carrolstress}
    T^\mu{}_\nu=\Big(\begin{matrix}
        P & 0 \\ -\frac{\tilde{\cal E}+ P}{v} & -\tilde{\cal E}
    \end{matrix}\Big)\ ,
\end{equation}
where $\tilde{\cal E}={\cal E}-\vec v\cdot {\vec P}_{mom}$ as before. To obtain this form, we have used that ${\cal E}+P=0$ as this is a general feature for Carroll fluids with $v\neq 0$, independent of conformal symmetry \cite{deBoer:2017ing}. If we do impose conformal symmetry, requiring the vanishing of the trace, we get the relation $\tilde {\cal E}=P$, and then \eqref{Carrolstress}
reduces to \eqref{CarrollstressCFT}, as it should.\\

\textbf{The first law}\\

Something interesting happens with the first law of thermodynamics. As observed already in e.g. \cite{Barnich:2012xq,Bagchi:2012xr,Bagchi:2013lma}, the first law for the cosmological solution comes with the wrong sign in front of the $T{\rm d}S$ term. Indeed, collecting all previous formulae, one can verify (with $J\equiv J_L$)
\begin{equation}
    {\rm d}(M\hbar c)=-T{\rm d}S-\hbar \omega {\rm d}J\ ,
\end{equation}
with frequency $\omega=\Omega/R$. From the point of view of the bulk, the minus sign in front of $T{\rm d}S$ has its origin in the fact that it comes from the flat space limit of inner horizon thermodynamics. From the point of view of the boundary, the minus sign is actually not there. The reason is that the energy density of the dual Carroll system is negative with ${\cal E}=-M\hbar c$, and therefore we have on the boundary the usual thermodynamic first law
\begin{equation}
    {\rm d}{\cal E}=T{\rm d}S-v\, {\rm d}P_{mom}\ ,
\end{equation}
where we used $\omega=v/R$ and $P_{mom}=-\hbar J/R$ as derived above. All this is consistent with the general relations found in \cite{deBoer:2017ing}. The first law of Carroll thermodynamics comes with the standard plus sign, but the specialty is that the Carroll energy density is typically negative because of ${\cal E}=-P$. This is a generic feature of Carroll systems. The example of the quantum mechanical system in the previous section also showed this, as we found there that ${\cal E}=-k_BTN$. In the next section, we show this as well for higher dimensional free field theories.

\section{Free scalar in $d$ dimensions}

\par In this section we discuss the Carroll limit of a free scalar field in $d$ spatial dimensions.  The aim is again to study the partition function and the Carroll limit of various thermodynamic quantities. Contrary to the previous section, we have no immediate application to holography here, but instead we focus on the procedure of Carroll limits with imaginary chemical potentials. 

We will study the theory on $\mathbb{R}^{d}\times S^1_R$ with a chemical potential $v$ in the direction of the circle. After Wick rotating to imaginary time, we have a Euclidean theory on $\mathbb{R}^{d-1}\times T^2$ at finite temperature. Partition functions for massless scalar fields were computed and analyzed in \cite{Alessio:2021krn}, so we can simply use their results as a starting point. The partition function can be written as an expansion in modified Bessel functions of the second kind, see eqn. (3.53) in ref \cite{Alessio:2021krn} (with $L_d=2\pi R$),
\begin{align}\label{lnZd}
    \ln Z&=-\beta E_C^d+
    \frac{V_{d-1}\Gamma(d-1)\zeta(d)}{2^{d-2}\pi^{\frac{d-1}{2}}\Gamma(\frac{d-1}{2})(\beta\hbar c)^{d-1}}\nonumber\\
    &+2\frac{V_{d-1}}{(2\pi R)^{d/2}(\beta\hbar c)^{\frac{d-2}{2}}}    
    \sum_{n_d\in\mathbb{Z}^*}\sum_{m\in\mathbb{N}^*}\Big(\frac{|n_d|}{m}\Big)^{d/2}K_{\frac{d}{2}}\Big( m|n_d|\frac{\beta\hbar c}{R}\Big)e^{ im|n_d|\frac{\beta\hbar v}{R}}\hspace{0.2cm}.
\end{align}
Here the volume is $V_{d-1}=\prod_{I=1}^{d-1}L_I$ and $R$ is the radius of $S^1$. The Casimir energy $E_C^d$ can be evaluated using zeta function regularization in $\mathbb{R}^{d-1}\times T^2$ to the following
\begin{equation}
    E_C^d=-\frac{V_{d-1}\hbar c}{(2\pi R)^d}\frac{\Gamma\big(\frac{d+1}{2}\big)\zeta(d+1)}{\pi^{\frac{d+1}{2}}}\ .
\end{equation}
For $d=1$, it yields the correct result for 2d CFTs. The Casimir energy is independent of the chemical potential. $v$ only shows up in the exponentials that can also be written as $e^{im|n_d|v/v_{th}}$ and which has manifest symmetry $v\to v+2\pi v_{th}$, just as in the previous sections.

Just as for 2d CFTs, the low temperature behavior is determined by the Casimir energy. Notice again that it is linear in the speed of light $c$, independent of the number of dimensions. In the Carroll limit, one would get a vanishing result, but if we repeat the exercises for $N$ scalars, we can get a non-zero and finite answer for the Casimir in the Carroll limit if we take $cN$ to be fixed.

The large temperature limit of \eqref{lnZd} is hard to extract, since many terms in this expression are blowing up as $\beta\to 0$. The way to deal with it, similar to 2d CFTs, is to use modular transformation properties. Since the theory is formulated on a torus, one can introduce the same modular parameter as for 2d CFT's, 
\begin{equation}\label{taumodpar2}
  \tau\equiv\tau_1+i\tau_2=\frac{\beta v\hbar}{2\pi R}+i\frac{\beta c \hbar}{2\pi  R}=\frac{v}{2\pi v_{th}}+i\frac{c}{2\pi v_{th}}\ ,\qquad  v_{th}=\frac{R}{\beta \hbar}\ ,
\end{equation}
such that we can write the Casimir energy as
\begin{equation}\label{ECtau2}
    -\beta E_C=\frac{V_{d-1}\xi_{d+1}}{(2\pi R)^{d-1}}\tau_2\ ,\qquad \xi_{d+1}\equiv \frac{\Gamma\big(\frac{d+1}{2}\big)\zeta(d+1)}{\pi^{\frac{d+1}{2}}}
\end{equation}
We can then study the behavior of the partition function under modular transformations. This was in fact also done in \cite{Alessio:2021krn}, and earlier in \cite{Shaghoulian:2015kta}, where it was shown that under a general modular transformation
\begin{equation}
    \tau'=\frac{a\tau+b}{e\tau+f}\ ,
\end{equation}
the logarithm of the partition function transforms as
\begin{equation}
    \ln Z'=|(e\tau+f)^{d-1}|\ln Z .
\end{equation}
Under the $S$-generator $\tau\to \tau'=-1/\tau$, we have $ 
    \ln Z (\tau', \bar\tau')=|\tau|^{d-1}\ln Z(\tau,\bar \tau)$ and
by using transforming the expression for $E^d_C$ as in \eqref{ECtau2}, one finds
\begin{align}
    \ln Z&=\frac{V_{d}\,c\xi_{d+1}}{\beta^d\hbar^d}\frac{1}{\big(c^2+v^2\big)^{\frac{d+1}{2}}}+\cdots\ ,
    \end{align}
where the dots stand for subleading terms that are not extensive in the total volume $V_d=2\pi RV_{d-1}$. Notice that $\ln Z$ vanishes in the strict Carroll limit, but we will consider the leading, nonzero term, in the small $c/v$ expansion.

\par We now derive the thermodynamic quantities for arbitrary dimensions. We follow the same prescription as used in section 3.1. Namely, we compute the generalized energy, average momentum and pressure, and find
\begin{align}
    \tilde{E}&=E-ivP_{mom}=-\frac{\partial\ln Z}{\partial \beta}=\frac{dV_{d}\,c\xi_{d+1}}{\beta^{d+1}\hbar^d}\frac{1}{\big(v^2+c^2\big)^\frac{d+1}{2}}    \ ,\\
    P_{mom}&=\frac{1}{i\beta}\frac{\partial\ln Z}{\partial v}=\frac{i}{\beta}\frac{(d+1)V_{d}\,c\xi_{d+1}}{\hbar^d\beta^d}\frac{v}{\big(c^2+v^2\big)^{\frac{d+3}{2}}}\hspace{0.2cm},\\
    P_{pr}&=\frac{1}{2\pi\beta V_{d-1}}\frac{\partial\ln Z}{\partial R}=\frac{c\xi_{d+1}}{\hbar^d\beta^{d+1}}\frac{1}{\big(v^2+c^2\big)^\frac{d+1}{2}}\hspace{0.2cm}.
\end{align}
We notice again the relation $\tilde{\cal E}=dP_{pr}$ as it should for conformal matter, as for $v=0$, the energy density is ${\cal E}=\tilde{\cal E}$. For non-zero $v$, the energy density is 
\begin{equation}
    \mathcal{E}=\frac{c\xi_{d+1}}{\hbar^d\beta^{d+1}}\frac{dc^2-v^2}{\big(v^2+c^2\big)^{\frac{d+3}{2}}}\hspace{0.2cm} .
\end{equation}
It becomes negative again for $v$ larger than the critical velocity $v_{c}=\sqrt{d}c$. In the Carroll limit, we get
\begin{equation}
    {\cal E}\to {\cal E}_C=-\frac{c\xi_{d+1}}{\hbar^d\beta^{d+1}}\frac{1}{v^{d+1}}\ .
\end{equation}
From these quantities it is straightforward to deduce the equation of state. The computation leads to the following formula
\begin{equation}
    \frac{\mathcal{E}}{P_{pr}}=\frac{dc^2-v^2}{c^2+v^2}\hspace{0.2cm},
\end{equation}
which is exactly the same relation as for the quantum mechanical system of massless relativistic particles. In the Carroll limit, we then get again an equation of state with $w=-1$, hence ${\cal E}+P_{pr}=0$.

Finally the entropy can be computed to be, with $D=d+1$ the number of spacetime dimensions,
\begin{equation}
    \frac{S}{k_B}=\frac{D\xi_DV_d}{(\beta \hbar c)^{D-1}}\frac{1}{(1+\frac{v^2}{c^2})^{D/2}}\ ,
\end{equation}
which generalizes \eqref{CardyEntropy} from $D=2$ to higher $D$. In the Carroll limit, for $v=0$, this becomes,
\begin{equation}\label{divEntv0}
    \frac{S_C}{k_B}=\frac{D\xi_DV_d}{(\beta \hbar c)^{D-1}}=D\xi_D\Big(\frac{v_{th}}{c}\Big)^{D-1}\ ,
\end{equation}
which diverges in the strict Carroll limit $v_{th}\gg c$. For $v\neq0$, we find a different behavior, as the Carroll limit $v\gg c$ yields
\begin{equation}
    \frac{S_C}{k_B}=\frac{D\,c\xi_DV_d}{\hbar^{D-1}}\frac{(k_BT)^{D-1}}{v^D} \ .
\end{equation}
In the strict Caroll limit, this vanishes, but when we add $N$ fields, we can keep this expression finite in the large $N$ limit if we keep $cN$ fixed. 

\section{Outlook}

We have determined the behavior of partition functions of various conformal systems in the Carroll regime.  For 2d holographic CFTs, there is an interesting and universal story that connects the Carroll limit to the flat space limit of BTZ black holes. We have shown this by taking the Carroll limit on the Cardy regime of a relativistic CFT$_2$, but not by first taking the Carroll limit and then going to high temperatures. There is still an open question or whether there is an order of limits issue. When the partition function in the Carroll regime obeys the modular property \eqref{CarrollModular}, we expect there to be no issue, but we don't know if all Carroll systems obey this modularity.

One may also wonder if there is a higher dimensional version of our analysis. One immediate difference is that, contrary to asymptotically flat 3d gravity in which there are no black holes, we do have black holes in higher dimensions. The situation in AdS$_4$ and in higher dimensions, is that there are large and small black holes. Large black holes have outer horizons larger than the AdS radius, $r_+\gg R$, and hence, similarly to BTZ, the outer horizon of the large black hole blows up to the null boundary on which the putative Carroll dual field theory lives. The divergence of the partition function and entropy mentioned in the introduction then has a physical origin, namely it is caused by the divergence of the entropy of large AdS-black holes in the flat space limit.

Small black holes in AdS don't feel the AdS curvature; they survive in the flat space limit and become black holes in asymptotically flat space. Let's illustrate this with some known formulas (now in natural units) for the simple example of a Schwarzschild black hole in AdS$_4$. Since there is no rotation, we may simply set $v=0$ in the dual field theory. The horizon is determined by the zero of (there is only one real root)
\begin{equation}
    f(r_+)=1-\frac{2M}{r_+}+\frac{r^2_+}{R^2}=0\ .
\end{equation}
The entropy and temperature are given by
\begin{equation}
    S=\pi r_+^2\ ,\qquad T=\frac{1+\frac{3r_+^2}{R^2}}{4\pi r_+}\ .
\end{equation}
For large black holes, $r_+\gg R$, we find the approximate formula for the root $r_+\approx (2MR^2)^{1/3}$, which, keeping $M$ fixed, blows up to the null boundary in the flat space limit $R\to \infty$. The entropy of such a large black hole diverges since it grows like $S\propto R^4 T^2$. Notice such a quadratic dependence $S\propto T^2$ as well for the free scalar field, see \eqref{divEntv0}, which diverges as well in the strict Carroll limit. While free scalars are not not dual to the Schwarzschild AdS black hole, we expect the same scaling relations to hold. For small black holes, $r_+ \ll R$ and the horizon becomes the location of a Schwarzschild black hole in the flat space limit, $r_+=2M$. The relation between entropy and temperature now becomes $S=1/16\pi T^2$. Producing this from a microscopic Carroll boundary action is of course an open problem as small black holes are already subdominant in the AdS partition function before taking the flat space limit.

\section*{Acknowledgements}

It is a pleasure to thank Jan de Boer, Troels Harmark, Emil Have and Niels Obers and Dirk Schuricht for stimulating discussions.

\bibliographystyle{JHEP}
\bibliography{ref}

\end{document}